\documentclass[10pt]{article}
\usepackage{latexsym,graphicx}
\usepackage[left=2.5cm,top=2.5cm,right=2.5cm,bottom=1.5cm]{geometry}
\usepackage{textcomp}
\usepackage{amsmath}
\usepackage{amssymb}
\usepackage{graphicx}
\begin{document}
    \begin{center}
        \Large{\bf{A Dark Energy Quintessence  Model of the Universe  }} \\
        \vspace{2mm}
        \normalsize{G. K. Goswami$^1$, Anirudh Pradhan$^2$, A. Beesham$^3$ }\\
        \vspace{2mm}
        \normalsize{$^1$ Department of Mathematics, Kalyan P. G. College, Bhilai-490 006, C. G., India}\\
        \vspace{2mm}
        \normalsize{$^1$ Email: gk.goswami9@gmail.com} \\
        \vspace{2mm}
        \normalsize{$^2$ Department of Mathematics, Institute of Applied Sciences and Humanities, G L A University,
            Mathura-281 406, Uttar Pradesh, India  \\
            \vspace{2mm}
            E-mail: pradhan.anirudh@gmail.com} \\
        \vspace{2mm}
        \normalsize{$^3$ Department of Mathematical Sciences, University of Zululand, Kwa-Dlangezwa 3886, South Africa  \\
            \vspace{2mm}
            E-mail: beeshama@unizulu.ac.za}
    \end{center}
\begin{abstract}
In this paper, we have presented a model of the FLRW universe filled with matter
and dark energy fluids, by assuming an ansatz that deceleration parameter is a
linear function of the Hubble constant. This results in a time-dependent DP
having decelerating-accelerating transition phase of the universe. This is a
quintessence model $\omega_{(de)}\geq -1$. The quintessence phase remains
for the period $(0 \leq z \leq 0.5806)$. The model is shown to satisfy current
observational constraints. Various cosmological parameters relating to the
history of the universe have been investigated.
\end{abstract}
\section{Introduction}
We begin  with the famous quote by  Allan Sandage \cite{ref1} that ``All of observational cosmology is the search for two numbers: 
Hubble (HP)  and deceleration parameters(DP) $H_0$ and $q_0$." Universe is a dynamical system in which its constituents (galaxies) 
travel like a discipline march of soldiers and move away from each other with Hubble's rate. Cosmological Principle (CP) is the
  basis of any model describing cosmology. As it is well-known \cite{ref2,ref3} that age and distance problems indicate that the 
  universe is  now accelerating which means that DP $q_0$ is negative. In the present scenario higher derivatives of scale factor such 
  as jerk parameter $j_0$, $s_0$ and $l_0$ do play role. It is concluded that  dark energy (DE) \cite{ref4} $-$ \cite{ref12} with
   negative pressure prevails all over the universe which is responsible for the said acceleration. Many researchers (Huterer, Turner, 
   Weller, Al-brecht, Polarski, Linder, Padmanabhan, Corasaniti and Alam ) \cite{ref13} $-$ \cite{ref20} have developed models of 
   the universe in which dark energy is taken as a perfect fluid with variable equation of state(EoS) parameter $\omega_{de}$ producing 
   negative pressure. These authors have considered different parametric forms of $\omega_{de}$. Off late, many authors 
   \cite{ref21} $-$ \cite{ref26} and references therein also developed DE perfect fluid models.\\
   
  The discovery of $2.73 K$ isotropic cosmic microwave background radiation (CMBR) motivated many authors to investigate 
FRW model with a two-fluid sources \cite{refM1}. Thermodynamics and two-fluid cosmological models consistent with current 
observations has been invistigated by Coley \cite{refM2}. Gromov {\it et al.} \cite{refM3} have studied the general 
properties of two-fluid FLRW models with energy transfer between dark energy and matter, independently of the specific 
mechanism of interaction. Interacting and non-interacting cases in non-flat FRW Universe in viscous dark tachyon cosmology have been 
studied by Setare {\it et al.} \cite{refN1}. Jamil {\it et al.} \cite{refN2} have discussed thermodynamics of dark energy 
interacting with dark matter and radiation. Singh and Chaubey \cite{refN3,refN4} have investigated the interacting 
two-fluid scenario for dark energy in anisotropic Bianchi type space-time. Reddy and Kumar \cite{refN5} have discussed 
the two-fluid scenario for dark energy model in scaler tensor theory of gravitation. Recently Amirhashchi {\it et al.} \cite{ref21,ref22}, 
Pradhan {\it et al.} \cite{ref23} and Saha {\it et al.} \cite{ref24} have studied interacting and non-interacting two-fluid-scenarios 
for dark energy models in FRW universe having constant deceleration parameter (DP) or negative DP. Recently, Garg {\it et al.} \cite{refN6}
have investigated transit cosmological models in FRW universe in interacting and non-interacting two fluid scenario. 
Goswami {\it et al.} \cite{refN7} have very recently focused on FRW accelerating universe with interacting dark energy.
Recently, Dagwal and Pawar \cite{refN8} have studied 
two fluids cosmological models with $G$ and $\Lambda$ in general relativity. Homogeneous and isotropic two-fluid cosmological models
are important because in these models two seperate fluids act as the source of the gravitational field as represented by the FRW 
space-time. The novelty of these two-fluid FRW models is that, using this approach, we can develop two separate models (i) in which the 
two-fluid do not interact with each other i.e., there is no possibility of energy transfer between the two matter 
sources (dark energy \& usual baryonic matter) and (ii) when the two fuilds interact with each other which allows 
the energy transfer between the two matter sources.\\

In this paper, we have presented a model of the FLRW (Friedman Lemaitre Robertson Walker) universe filled with matter and DE 
fluids, by assuming an ansatz that DP is a linear function of the Hubble constant. This results in a time-dependent DP having 
decelerating-accelerating transition phase of universe. This is a quintessence model  $\omega_{(de)}\geq -1$. The
quintessence phase remains for the period $(0 \leq z \leq 0.5806)$. The model is shown to satisfy current observational constraints.
Various cosmological parameters relating to the history of the universe have been investigated. \\

 The paper is structured as follows: In Sec. $2$, we set the initial field equations. In Sec. $3$, we described solution and physical 
 properties of DE model. Finally, Sec. $4$ is devoted to conclusions.
\section{Field equations:}

The Einstein field equations (EFEs) are given by
\begin{equation}
\label{eq1}
R_{ij}-\frac{1}{2}Rg_{ij} =-\frac{8\pi G}{c^{4}}T_{ij},
\end{equation}

where $R_{ij}$ is the Ricci tensor, $R$ the scalar curvature, and $T_{ij}$ the
stress-energy tensor. It is taken as $ T_{ij} = T_{ij}(m)+T_{ij}(de),$ where
$T_{ij}(m)=\left(\rho_m + p_m\right)u_{i}u_{j}-p_m g_{ij}$ and
$T_{ij}(de)=\left(\rho_{de}+p_{de}\right)u_{i}u_{j}-p_{de} g_{ij}.$
We take $u^{\alpha}=0; \; \; \alpha=1,2,3$.\\

The FLRW space-time (in units $c = 1$) is given by
\begin{equation}
\label{eq2}
ds{}^{2}=dt{}^{2}-a(t){}^{2}\left[\frac{dr{}^{2}}{(1+kr^{2})}+r^{2}({d\theta{}^{2}+sin{}^{2}\theta
    d\phi{}^{2}})\right],
\end{equation}
where $a(t)$ stands for the scale factor and $k$ is the curvature parameter, $k =
+1$ for closed , $k = -1$ for open and $k = 0$ for  a spatially flat universe.
Solving the EFEs (\ref{eq1}) for the FLRW metric (\ref{eq2}), we obtain the
following system of equations: $2\frac{\ddot{a}}{a}+H^{2} = -8\pi G
(p_m+p_{de}) + \frac{k}{a^{2}}$ and $H^{2} =  \frac{8\pi G}{3}(\rho_m+\rho_{de}) +
\frac{k}{a^{2}}$, where $H=\frac{\dot{a}}{a}$ is the Hubble constant. Here an over
dot means differentiation with respect to cosmological time $t$. We have
deliberately put the curvature term on the right of EFEs, as this term is made to
acts like an energy  term. For this, we assume that  the density and pressure
for the curvature energy are as follows $\rho_{k}=\frac{3 k}{8\pi Ga^{2}}, ~~
p_{k}=-\frac{k}{8\pi Ga^{2}}$. Finally, we get following equations for  FLRW
cosmology
\begin{equation}
\label{eq3}
2\frac{\ddot{a}}{a} + H^{2} = -8\pi G p
\end{equation}
\begin{equation}
\label{eq4}
H^{2}=\frac{8\pi G}{3}\, \rho,
\end{equation}
where $\rho = \rho_m + \rho_{de}+\rho_{k}$ and $p =p_m+p_{de}+p_{k}$ are the
total energy and pressure of the universe. The energy conservation law
$T^{ij}_{;j}=0$ yields
\begin{equation}
\label{eq5}
\dot{\rho}+3H(p+\rho)=0.
\end{equation}
As $\dot{\rho_{k}}+3H(p_{k}+\rho_{k})=0,$ so $\frac{d}{dt}{(\rho_{m}+\rho_{de})}+3H(p_{m}+p_{de}+\rho_{m}+\rho_{de})=0.$
We assume that both matter and dark energies are minimally coupled so that they are conserved simultaneously, i.e. 
$\;\dot{\rho_{m}}+3H(p_{m}+\rho_{m})=0,\;\;\dot{\rho_{de}}+3H(p_{de}+\rho_{de})=0.$ At present matter content in the universe
 is in form of dust for which $p_m=0$. We assume EoS for dark energy as $p_{de} = \omega_{de} \rho_{de}$. Integration of energy 
 conservation laws yields
 \begin{equation}
 \label{eq6}
 \rho_{m} = (\rho_{m}) _0 (1+z)^3,\;\; \rho_{de} = (\rho_{de}) _0~ exp \left( 3\int^z_0 \frac{(1+\omega_{de})dz}{1+z}\right),
 \end{equation}
  where we have used $\frac{a_0}{a} = 1+z.$
  The EoS for the curvature energy is obtained as $p_{k}=\omega_{k}\rho_{k}~\mbox{where}~ \omega_{k}=-1/3.$ This gives
  $\rho_{k}\varpropto a^{-2}= (\rho_{k})_{0}\left[\frac{a_0}{a}\right]^2.$ We define density parameters for matter, DE and curvature as
  $\Omega_{m}=\frac{\rho_{m}}{\rho_{c}},\Omega_{de}=\frac{\rho_{de}}{\rho_{c}}~~\&~~\Omega_{k} =\frac{\rho_{k}}{\rho_{c}}$
  where $ \rho_{c}=\frac{3H^{2}}{8\pi G},$ is the critical density. Therefore the FLRW field equations change to
  \begin{equation}
  \label{eq7}
  H^2(1-\Omega_{de})=H^{2}_{0}\left[(\Omega_{m})_{0} (1+z)^3+(\Omega_{k})_{0}
  (1+z)^{2}\right],
  \end{equation}
  and
  \begin{equation}
  \label{eq8}
  2q = 1 + 3\omega_{de}\Omega_{de}+ 3\frac{H^2_0}{ H^2}
  \omega_{k}(\Omega_{k})_{0} (1+z)^2 ,
  \end{equation}
  where $q$ is the DP defined by $q=-\frac{\ddot{a}}{aH^2}.$
  \section{Results and discussion:}
  In the above, we have found two field equations (\ref{eq7}) and (\ref{eq8}) in five unknown variables $a,~H,~q,\Omega_{de}$
  and $ ~\omega_{de}$. Therefore, for a complete solution, we need three more  relations involving these variables.
   As it has been discussed in the introduction that in view of the recent observations of Type Ia  supernova [01-12] ,
   there is a need of a time-dependent DP which describe decelerated expansion in the past($z\geq 1$) and accelerating
  expansion at present, so there must be a transition from deceleration to acceleration. DP must show the 
  change in signature. So we must use an ansatz to determine proper DP. Earlier Akarsu et al. \cite{ref27} used hybrid expansion
   law [HEL] to express deceleration to acceleration transition and cosmic history.
   Before this, so many parametrization of $\omega_{de}$ were proposed
   such as Linear-red shift parametrization \cite{ref36,ref37}, Chevallier-Polarski-Linder
   parametrization \cite{ref38,ref39}, Jassal-Bagla-Padmanabhan parametrization \cite{ref40,ref41} ,
   Upadhye-Ishak-Steinhardt parametrization \cite{ref42}, Hannestad-M¨ortsell
  parametrization \cite{ref43}, Lee parametrization \cite{ref44} etc. \\
  
  Experiment and observation are the final arbiter of all theoretical speculation. The crowing glory of any theory is its ultimate 
  verification by experiment and observation. In order to get an expanding model of the universe consistent with observations, one needs 
  a Hubble parameter $H$ such that the model starts with a deceleration expansion followed by an accelerating expansion at late time. 
  Following the lines floated by Ellis and Madsen \cite{refN9}, we choose a functional form of $H$, which despite quite simple, will 
  describe both decelerating and accelerating phase of the universe $H(a) = \ell(a^{-n} + 1)$, where $\ell$ and $n$ are positive constants. 
  Banerjee and Das \cite{refN10} investigated a quintessence model with minimally coupled scalar field by choosing a particular 
  form of the DP $q$. This cosmological scenario is in good agreement with recent supernovae observations \cite{ref2}-\cite{ref7}. 
  Sing \cite{refN11,refN12} has also investigated Bianchi type-$I$ and $V$ cosmology by considering one-parameter function 
  $H(a) = \ell(a^{-n} + 1)$.  \\
  
 Motivated by above investigations, in this paper we consider $q$ as a linear function of the Hubble parameter which was earlier
used by \cite{ref10,ref28,refN13,refN14,refN15,refN16} in different context of cosmological models.
    \begin{equation}
    \label{eq9}
    q  = \alpha H + \beta
    \end{equation}
    Here $\alpha$, and $\beta$ are arbitrary constants. This behaviour of $q$ presents a unified description of the evolution of 
    the universe which starts with a decelerating expansion and expands with acceleration at late time.\\  
    
    The jerk parameter $j= \frac{\dddot{a}}{aH^3}$ is related to $\dot{q}$ through
   \begin{equation}
   \label{eq10}
    \dot{q} = (q(2q+1)- j) H
   \end{equation}
   From Eqs. (\ref{eq9}), (\ref{eq10}) and $\dot{H} = -(1+q)H^2$, ~ we get $\alpha  = \left
   (j - q (2 q + 1))/(1 + q)H\right)_{z=0} $ and $\beta =\left( (-j + q (2 + 3 q))/(1 + q)\right)_{z=0}$
   Now, we present followings data's regarding values of cosmological parameters at present due to Amirhashchi and Amirhashchi \cite{ref29}
   $H_0=69.9\pm1.7,~ \Omega_{m0}=0.279_{-0.016}^{+0.014},~ \Omega_{de0}=0.721_{-0.014}^{+0.016},~ j_0 = 1.038_{-0.035}^{+0.061}$ and
     $z_t = 0.707 \pm 0.034.$  They have estimated these values using Pantheom compilation \cite{ref30} and JLA data set \cite{ref31}
    These data's enable us to determine constants $\alpha$ and $\beta$. We get  $\alpha = 31.619, \beta = -2.763.$ Eq. (9)
    provides following differential equation and its solution.
   \begin{equation}
   \label{eq11}
      (1 + z) H_z = (\delta + \alpha H ) H ,~ \delta= 1+\beta
   \end{equation}
  \begin{equation}
  \label{eq12}
   H = (A \delta (1 + z)^\delta)/(1 - A \alpha (1 + z)^\delta),
  \end{equation}
  where $A$ is constant of integration. It is determined  by applying initial condition that $ H_0= 0.07~ Gyr^{-1} $ as
  $ A = 7./(100~\delta + 7~\alpha ) = 0.155545.$ We have also used $\dot{z}=-(1+z)H.$
  The transition red shift $z_t$ where DP is zero, is obtained as
 \begin{equation}
 \label{eq13}
  z_{t}= \left(\frac{-A y}{x}\right)^{\frac{1}{y+1}}-1 = 0.386719 \simeq  9.4699 Gyr
 \end{equation}
 This means as from now, before $4.25$ Gyr, universe  started accelerating. It is interesting to note that
 transition red shift obtained by us is matching with the work referred in the introduction [13]-[20].
 We present following plots to show our results. \\
 
  \begin{figure}[ht]
    \centering
    \includegraphics[width=10cm,height=5cm,angle=0]{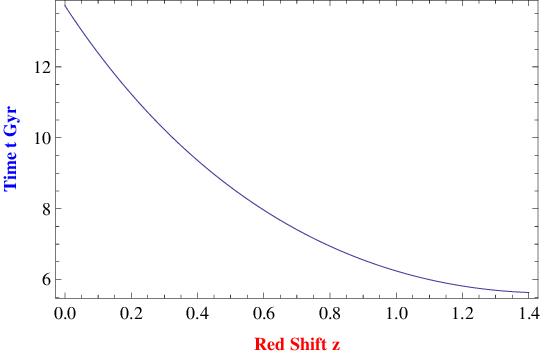}
    \caption{ Time ($t$) versus red shift($z$) Plot. This is drawn by solving  $\dot{z}=-(1+z)H$ and Eq. (\ref{eq12})}. 
 \end{figure}
 \begin{figure}[ht]
    \centering
    \includegraphics[width=10cm,height=5cm,angle=0]{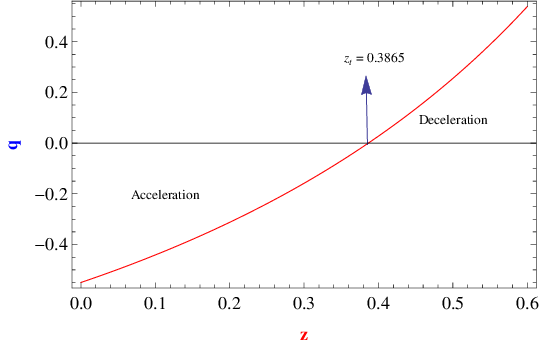}
    \caption{Variation of $q$ with $z$. This is drawn by solving  $\dot{H} = -(1+q)H^2$, $\dot{z}=-(1+z)H$ and Eq. (\ref{eq12}).
    Transition red shit is $ z_t $= 0.3865. }
 \end{figure}

\textbf{(i) $\chi^2$ for Hubble Parameter:}\\

We present following table which describe observed values of Hubble parameter
along with error and corresponding theoretical results obtained as per our model in the range ($0\leq z \leq1$).
 \begin{center}
    Table-1
 \end{center}

$\begin{array}{|c|c|}
 \hline
 {\bf{z}} ~ = ~0.07,~0.1,~0.12,~0.17,~0.179,~0.199,~0.2,~0.27,~0.35,~0.352,~0.4,~0.44,~0.48,
~0.57,\\~0.593,~0.6,~0.781,~0.875,~0.88,~0.9,~1.037\\
\hline
 {\bf{H_{ob}}}
~=~  0.0746,~ 0.0787,~ 0.0908,~ 0.0846,~ 0.0848,~ 0.0971,~ 0.0844,~ 0.0991,~ 0.1064,~ 0.0899,\\
  0.0934,~0.0989,~0.1073,~0.1278,~0.092,~0.1196,~0.1636,~0.181,~0.1432,~0.2066,~0.1907\\
 \hline
{\bf{H_{er}}}
 ~=~~0.0302,~0.0143,~0.0374,~0.0085,~0.0143,~0.0173,~0.0079,~0.0634,~0.0132,~0.0062,\\
~0.0081,~0.0071,~0.0122,~0.0173,~0.0409,~0.0235,~0.0343,~0.0184,~0.0143,~0.0409,~0.0515\\
 \hline
{\bf{H_{th}}}
~=~0.0723,~0.0734,~0.0741,~0.0762,~0.0765,~0.0774,~0.0774,~0.0808,~0.0851,~
 0.0852,~\\ 0.0882,~0.0909,~0.0931,~0.1014,~0.1036,~0.1043,~0.1274,~0.1452
 ,~0.1463,~0.1509,~0.1942\\
 \hline
 \end{array}$\\
\\

The observed values and error are taken from a latest paper by Farook {\it at el.} \cite{ref32}. The Hubble data's are converted
into $Gyr^{-1}$ unit. In order to get quantitative closeness of theory and observation, we obtain $\chi^2$ from the following formula
$$\chi^2= \frac{(H_{ob}-H_{th})^2}{H_{er}^2}$$
It comes to $19.41469$ which is $92.45\%$ over 21 data's, which shows best fit in theory and observation.
We present following fig. to display this closeness.\\

 \begin{figure}[ht]
    \centering
    \includegraphics[width=10cm,height=5cm,angle=0]{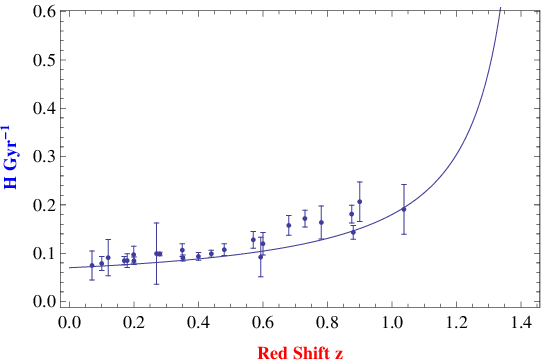}
    \caption{Error bar plot of Hubble parameter $H$ showing proximity of theory and observation.}
    \end{figure}
    \textbf{(ii) DE Parameter $\Omega_{de}$ and EoS  $\omega_{de}$ }\\

    Now, from Eqs. (\ref{eq7})-(\ref{eq9}), the DE parameter $\Omega_{de}$ and EoS
    parameter $\omega_{de}$  are given by the following equations  and are solved numerically.
    \begin{equation}
    \label{eq14}
    H^2 \Omega_{de} = H^2 - (\Omega_{m})_0 H^2_0 (1+z)^{3}
    \end{equation}

    \begin{equation}
    \label{eq15}
    \omega_{de}=\frac{H^2(2\alpha H +2\beta-1)}{3[H^2-H_0^2(\Omega_{m})_0(1+z)^{3}]}.
    \end{equation}
    where we have taken  $(\Omega_{k})_0 = 0$ for the present spatially flat universe. We solve Eqs. (\ref{eq14}) and (\ref{eq15}) with 
    the help of Eq. (\ref{eq12}) and present the following figures $4$ and $5$ to illustrate the solution.
    
    \begin{figure}[ht]
        \centering
        \includegraphics[width=10cm,height=5cm,angle=0]{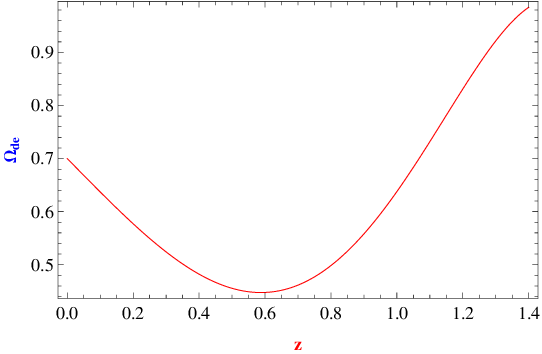}
        \caption{Plot of $\Omega_{de}$ versus red shift ($z$). This is drawn by solving  Eq. (\ref{eq14}) 
        and Eq. (\ref{eq12}). $H_0 = 0.07 Gyr^{-1}$ and  $\Omega_{m0}$=0.279 .}
    \end{figure}
    \begin{figure}[ht]
        \centering
        \includegraphics[width=10cm,height=5cm,angle=0]{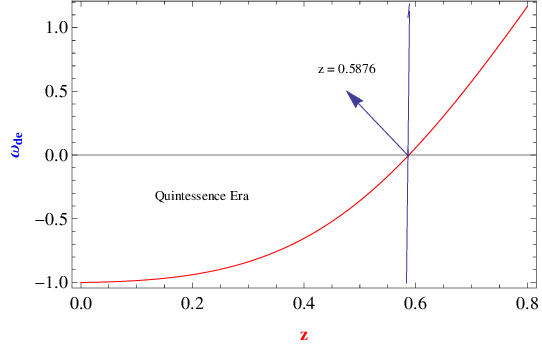}
        \caption{Plot of $\omega_{de}$ versus $z$. This is drawn by solving Eq. (\ref{eq15}) and Eq. (\ref{eq12}). 
        Figure shows that this is a quintessence model.} 
    \end{figure}
     The figures show that this is a quintessence model $\omega_{(de)}\geq -1$ which  remains for the period
    $(0 \leq z \leq 0.5806)$. DE favors deceleration at $z \geq 0.5806$ and will help in structure formation at $z \geq 4$.
    As per our model, the present value of DE parameter is $0.7$. It decreases over the past, attains a  minimum value
    $\Omega_{de} \sim 0.45$ at $z \sim 0.6$, and then it again increases with red shift. It is interesting to note that the growth
    of EoS parameter for DE $\Omega_{de}$ over red shift during quintessence regime
    is matching with the pioneering work by Steinhardt {\it et al.} \cite{ref33} and Johri \cite{ref34} \\

    We look carefully Fig. $4$ in context of Fig. $5$. In fact as per Fig. $5$, the transition red shift is $z_{te} = 0.5806$.
    As we expressed in our explanation, dark energy will begin its roll of opposing deceleration and favoring acceleration during
    $0\leq z\leq 0.5806$. Before, i.e., $z\geq 0.5806$, $\omega_{de}$ is positive, so dark energy favours deceleration. We may 
    say that the validity of Fig. $4$ is only during the said tenure $0\leq z\leq 0.5806$. During this DE always increases with time.
    This feature provides a possible way out for the coincidence problem affecting many quintessence models. Unfortunately this
    the scenario is still affected by the cosmological coincidence problem (CCP) \cite{refP1}.  \\
    
      \textbf{(iii) Luminosity Distance:}\\

     The red shift-luminosity distance relation \cite{ref35} is an important observational tool to study
     the evolution of the universe. The expression for the luminosity distance ($D_L$) is obtained in term of red shift 
     as the light coming out of a distant luminous body gets red shifted due to the expansion of the universe. We determine 
     the flux of a source with the help of luminosity distance. It is given as $D_{L}=a_{0} r (1+z).$ where r is the radial 
     coordinate of the source. For the FLRW metric Eq. (2), the radial co-ordinate $r$ is obtained as 
     $r = \int^r_{0}dr = \int^t_{0}\frac{cdt}{a(t)} = \frac{1}{a_{0}H_{0}}\int^z_0\frac{cdz}{h(z)},$
     where we have taken $k=0$ and have used $ dt=dz/\dot{z}, \dot{z}=-H(1+z)~\&~ h(z)=\frac{H}{H_0}.$ Therefore, we get the
     luminosity distance  as:
     \begin{eqnarray}
     \label{eq16}
    \frac{H_{0} D_{L}}{c} = (1+z)\int^z_0\frac{dz}{h(z)}.
     \end{eqnarray}

     \textbf{(iv) Distance modulus $\mu$ and Apparent Magnitude $m_{b}$:}\\

     The distance modulus $\mu$ \cite{ref12} is derived as
     \begin{equation}
     \label{eq17}
     \mu  = m_{b}-M = 5log_{10}\left(\frac{D_L}{Mpc}\right)+25 = 25+  5log_{10}\left[\frac{c(1+z)}{H_0}
     \int^z_0\frac{dz}{h(z)}\right]
     \end{equation}
     The absolute magnitude $M$ of a supernova \cite{ref12} is $ M=16.08-25+5log_{10}(H_{0}/.026c)$. Using this, the
     apparent magnitude $m_b$ is obtained as
     \begin{equation}
     \label{eq18}
     m_{b}=16.08+ 5log_{10}\left[\frac{1+z}{.026} \int^z_0\frac{dz}{h(z)}\right].
     \end{equation}
     We solve Eqs. (16)$-$ (18) with the help of Eq. (\ref{eq12}). Following is the value of the integral $$\int^z_0\frac{dz}{h(z)}
     \simeq (1.25 z +  0.09 (1- (1. + z)^{2.76})$$\\

     \textbf{(v) $\chi^2$ for Distance Modulus `$\mu$':}\\

      Our theoretical results have been compared with
     SNe Ia related   $581$ data's from Pantheon compilation \cite{ref30} with possible error in the range ($0\leq z \leq1.3$.) 
     and the derived model was found to be in good agreement with current observational constraints. In order to get quantitative
     closeness of theory and observation, we obtain $\chi^2$ from the following formula
     $$\chi^2= \frac{(\mu_ {ob}-\mu_{th} )^2}{\mu_{err}^2}$$ It comes to $562$ which is $96.73\%$ over $581$ data's, which shows 
     best fit in  theory and observation. The following figures $8$ depict the closeness of observational and theoretical results,
     thereby justifying our model.

     \begin{figure}[ht]
        \includegraphics[width=10cm,height=5cm,angle=0]{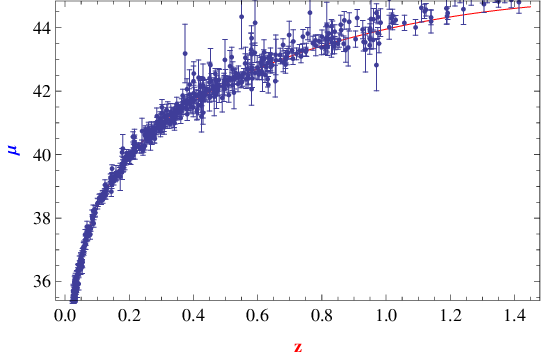}
        \caption{Plot of distance modulus ($\mu=M-m_b$) versus red shift ($z$). Crosses are SNe Ia related Pantheon compilation $581$ data's
         with possible corrections, $H_0$ = 0.07 $Gyr^{-1}$, $\Omega_{m0}$ = 0.279 and  $\Omega_{\Lambda0}$ = 0.721.}
   \end{figure}

       \section{Conclusion:}

         In the present paper, we have presented an FLRW universe filled with two fluids (barotropic and dark energy), by assuming
        DP as a linear function of the Hubble constant. This results in a time-dependent DP having a transition
        from past decelerating  to the present accelerating universe. The main findings of our model are itemized point-wise as follows.
        \begin{itemize}
            \item The expansion of the universe is governed by a  expansion law
             $q  = \alpha H + \beta,~ H= -\frac{A (\beta+1) (z+1)^{\beta+1}}{0.07 \left(A \alpha (z+1)^{\beta+1}-1\right)} $, where
            $ \alpha =  31.619, \beta = -2.76333$ and A = 0.15. This describes the transition from  deceleration to acceleration.

            \item The transition red shift is $z_t = 0.386$ and corresponding time is $T_t = 1.034$ Gigayear. The universe is now accelerating
             and it was decelerating before time $T_t$

            \item This is a quintessence model $\omega_{(de)}\geq -1$ which remains for the period
            $(0 \leq z \leq 0.5806)$. DE favors deceleration at $z \geq 0.5806$.
            
            \item the present value of DE parameter is $0.7$. It decreases over the past, attains a minimum value 
            $\Omega_{de} \sim 0.45$ at $z \sim 0.6$, and then it again increases with red shift.
            
            \item  The growth of EoS parameter for DE $\Omega_{de}$ over red shift during quintessence regime is matching with 
            the pioneer work by Steinhardt {\it et al.} \cite{ref33} and Johri \cite{ref34}.

            \item  In the present model dark energy doe not interact with a barotropic fluid. We have observed that for both cases, interacting
            and non-interactiong, the graphs come to be the same. It indicates that the dark energy component is so dominated that, its
            interaction with barotropic fluid does not affect the non-interacting scenario. So, we have only discussed the non-interacting
            model.
            
           \item It is to mention here that that the assumption of deceleration parameter $q$ as a function of $H$ is {\it adhoc} in the 
           sense that it does not result from known field theory. However, it does produce a solution that presents an appropriate description of
           the universe consistent with observations. The agreement with the observed universe is just qualitative.

            \item Our model is compatible with current observations. We have discussed it in different places with appropriate motivations 
            and their references.

            \item In Fig. $4$, we observe that $\Omega_{de}$ does not always increase. It is minimum at red shift $\approx 0.6$. If 
            we parametrize the expansion of the Universe in terms of red shift $z$, it appears a particular period of cosmic history 
            as density ration of DM and DE of order of one. This scenario at low value of $z$ suggests that the dominance of DE started 
            ``recently'' or even ``now'' on cosmological scale, giving rise to the CCP ``why now?'' \cite{refP1}.

        \end{itemize}

        In a nutshell, we believe that  proposed linear law may help in investigations of hidden matter like dark matter, dark energy 
        and black holes. \\
  
     \section*{Acknowledgement}
The authors (G. K. Goswami \&  A. Pradhan) sincerely acknowledge the Inter-University Centre for Astronomy and Astrophysics (IUCAA),
Pune, India for providing facilities where part of this work was completed during a visit. The authors are grateful to the anonymous reviewers
for constructive suggestions to make an improved version of the manuscript.


\end{document}